\documentclass[preprint,12pt,authoryear]{elsarticle}

\usepackage{amssymb}
\usepackage{amsmath}
\usepackage{xcolor}

\usepackage{algorithm}
\usepackage{algpseudocode}
\usepackage{graphicx} % Required for inserting images
\usepackage{caption}

\usepackage{anyfontsize}

\usepackage{subcaption}
\begin{document}

\begin{frontmatter}



\title{\bf Design an Algorithmic Stablecoin based on a Predictive PID Controller} %% Article title

\author[label1]{Yuqiao Zhao\corref{cor1}}\ead{yqzhao7@connect.hku.hk}
\author[label1]{Wai-Ki Ching}\ead{wching@hku.hk}
\author[label1]{Chi-Wing Wong}\ead{cwwongab@hku.hk}
\cortext[cor1]{Corresponding author}
\affiliation[label1]{organization={Department of Mathematics, The University of Hong Kong},%Department and Organization
            %addressline={}, 
            city={Hong Kong},
            postcode={999077}, 
            country={China}}

%% Abstract
\begin{abstract}
Cryptocurrencies issued based on blockchain technology have attracted increasing attention in recent years. 
However, their high volatility makes it difficult for them to become circulating currencies. This led to the creation of stablecoins. 
Existing stablecoins can be divided into three main categories: stablecoins pegged to fiat currencies, stablecoins collateralized by crypto assets, and algorithmically controlled stablecoins. 
Most existing stablecoins are either of the first two types. We design an algorithmic stablecoin controlled only by a predictive Proportional-Integral-Derivative (PID) controller, which uses the price forecasted by Autoregressive Integrated Moving Average (ARIMA) model as part of its input. Simulation experiments show that, under ideal market conditions, the stablecoin can achieve long-term goals that fluctuate within a pre-defined range. Furthermore, it consistently restores prices to the anchor interval in the presence of exogenous disturbances.
\end{abstract}

% \begin{highlights}
% \end{highlights}

%% Keywords

\end{frontmatter}



\section{Introduction}
\label{sec:introduction}
%% Labels are used to cross-reference an item using \ref command.
With the launch of Bitcoin in 2008, the concept of cryptocurrency gradually entered public consciousness and has experienced rapid development in recent years \citep{pernice2021cryptocurrency, angeloni202moneyInCrisis}. 
The establishment of the ``Crypto Strategic Reserve" in the United States in 2025 marks a significant step for cryptocurrencies into the financial system \citep{giordano2025cryptocurrencies}. As a parallel noncredit monetary system, token-based cryptocurrencies are different from digital currencies issued by central banks as liabilities \citep{giordano2025cryptocurrencies}. 
However, the violent volatility of the price of cryptocurrencies such as Bitcoin prevents them from serving the essential functions of traditional money \citep{angeloni202moneyInCrisis}. 
To address this issue, a type of cryptocurrency designed to maintain a stable value, known as a stablecoin is created. 

Stablecoins serve as a bridge between traditional finance and the cryptocurrency ecosystem, not only facilitating smoother fund transfers, but also mitigating the risks associated with the high volatility of other cryptocurrencies
\citep{lee2025stablecoinDepegging}. 
As listed in \citet{strohmeyer2024stablecoinCategories}, the existing stablecoins can be divided into two categories: 
custodial and non-custodial stablecoins. 
This categorization is determined by the presence of a central authority responsible for functions such as reserve management and token regulation. The first type of stablecoin, represented by Tether’s USDT \citep{tether2014usdtWhitepaper} and Circle’s USDC \citep{circle2018usdcWhitepaper}, is pegged to a reliable fiat currency, sacrificing decentralization in exchange for a relatively stable price. In contrast, the second type is regulated by smart contracts rather than central reserves, thus retaining the decentralized nature of cryptocurrencies. Crypto-backed stablecoins, using other cryptocurrencies as collateral, is a common practice of the second type. However, the high volatility of collateralized cryptocurrencies exposes the stablecoin to a higher risk of liquidation \citep{klages2022collateral}. For example, MakerDAO’s DAI \citep{maker2017daiWhitepaper} was backed only on Ethereum in the initial design. After experiencing a collateral shortfall caused by a deflationary deleveraging spiral, it included USDC as a new collateral \citep{klages2022collateral}. Founded on DAI, \citet{inoescu2020rai} applied the feedback controller to update the redemption price according to market price. Algorithmic stablecoins, which stabilize price through decentralized algorithms (smart contract) \citep{zhao2021algorithm}, is another sub-category of non-custodial stablecoins. 

Although most of the current non-custodial stablecoins are collateralized \citep{angeloni202moneyInCrisis}, algorithmic stablecoins are also being put forward continuously. Proposed by \citet{kereiakes2019terraWhitePaper}, TerraUSD (UST) utilized the seigniorage model to keep its price stable at \$1.  
TerraUSD has been the most successful while the worst algorithmic stablecoin, since it raised over \$60 billion capitalization in 15 months, but lost more than 90\% of its market capitalization within 7 days \citep{calandra2025terraCase}. 
Ampleforth \citep{kuo2019ampleforth} is a rebase-type stablecoin. The mechanism of it uses a 24-hour volume-weighted average price to actively determine supply adjustments, which directly shrinks or grows the number of coins held by users. Additionally, \citet{kazemian2020fraxV1} developed the first partial-collateral stablecoin Frax. A collateral ratio is dynamically determined to adjust the ratio of \$1 worth of collateral and Frax shares required to mint or redeem one Frax. 

\citet{akcin2024controlpayIEEE} observed that not only the current monetary polices, but also the token economies can be captured by control theory. However, to the best of our knowledge, few algorithmic stablecoins have been designed on the basis of control theory. The Proportional-Integral-Derivative (PID) control is a standard feedback-based loop mechanism to maintain a system variable at a setpoint \citep{visioli2006PIDSpringer}. As the survey results from \citet{samad2017controlSurvey} show, PID controllers are still the most widely adopted control law in industry, although other techniques are constantly being developed. Despite this, traditional PID controllers lack predictability and their heuristic and reactive features are insufficient to respond promptly to future situations \citep{akcin2024controlpayIEEE}. This led to the development of predictive PIDs, one type of it can be found in \citet{moradi2001predictivePID}, which is designed based on Generalised Predictive Controllers.

We are developing Conttsus, an algorithmic stablecoin that operates within a designated price corridor rather than adhering to a rigid peg. This flexible design creates valuable opportunities for arbitrageurs in the short term, while its overall stability ensures that it remains a reliable store of value for long-term investors. Similar to Ampleforth, we employ a price oracle to obtain real prices of the stablecoin. Then, a well-known prediction model Autoregressive Integrated Moving Average (ARIMA) \citep{box1970ARIMAHolden_Day} is applied to predict future prices. Then we employ a predictive PID model to regulate the supply of the stablecoin, which can correspondingly affect its price. Different from the rebase-type stablecoin, the supply to be adjusted here includes the amount of coins issued through the protocol and the available volume for market trading. 

The remainder of the paper is organized as follows. 
In Section \ref{sec:probState}, we reiterate the questions we are investigating, state the assumptions upon which our solutions are based, and explain the key symbols involved. 
Section \ref{sec:method} provides a thorough description of the prediction and control models and the method of conducting simulation experiments. 
Section \ref{sec:experiment} gives the initial settings and results of the simulation experiments. 
Ultimately, Section \ref{sec:conclusion} concludes the work and lists further research directions.

\section{Problem Statement}
\label{sec:probState}
Our objective is to design a stablecoin, Conttsus, that relies entirely on algorithms to maintain its stability. Unlike existing stablecoins whose prices are pegged to a certain value of currency (fiat currency or cryptocurrency), this stablecoin allows its price fluctuates within a given range. Its characteristics of long-term stability and short-term fluctuations can meet the needs of both arbitrageurs and savers. The algorithm design in our stablecoin protocols is based on the following assumptions. Firstly, the price of stablecoins is determined solely by supply and demand, which means we do not consider the impact of policy intervention or external shocks, etc. 
Secondly, we assume that all stablecoins are liquid and can be traded in arbitrarily small units. 

At time $t$, we denote the realized price sourced from the oracle as $p_t$ and the forecasted price as $\hat{p}_t$. 
The variables $Q^D_t$ and $Q^S_t$ represent the demand quantity and the quantity of circulating supply, respectively. Note that the circulating supply encompasses all tokens currently available for trading, inclusive of holdings by both the protocol and external investors. In reality, $p_t$, $Q^D_t$, 
and $Q^S_t$ are all non-negative. 
Upon the genesis of the protocol, the initial circulating supply of Conttsus will be set to $Q^S_0$. 
At the commencement of each timestamp $t$, the protocol executes an adjustment to $Q^S_t$. 
This is achieved by either restricting open market trading volume or expanding the token supply of the protocol, thereby ensuring the Conttsus price remains within the target corridor $[p^l,p^h]$ throughout normal market cycles. 
The process can be decomposed into two sub-problems. 
To start with, we need a prediction model to obtain $\hat{p}_t$. 
After that, a control model should be designed to decide the adjustment of 
$Q^S_t$ based on $\hat{p}_t$ and the historical price. 
Simulation experiments are designed to assess the feasibility of the process.

\section{Methodology}
\label{sec:method}

\begin{figure} [htbp]%% placement specifier
\centering
\includegraphics[trim=3cm 7.5cm 4.5cm 7.5cm, clip, width=1.0\textwidth]{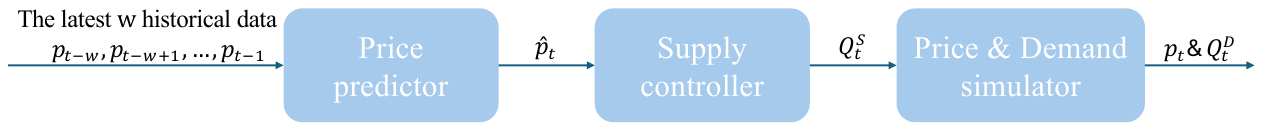}
\caption{Flowchart of the whole pipeline for the time interval $(t-1, t]$.}
\label{fig:flowchart}
\end{figure}
Figure \ref{fig:flowchart} displays the pipeline of stabilization mechanisms and simulation experiments in the time interval $(t-1, t]$. 
At the beginning of the time interval, the latest historical price data with size $w$ serve as a training set for the price predictor, which then are utilized to forecast $\hat{p}_t$. 
The predicted price together with all historical prices is used for the adjustment of the quantity supplied $Q^S_t$. 
Finally, at the end of the time interval, the price $p_t$ and the quantity demanded $Q^D_t$ are  calculated based on $Q^S_t$. 
The following subsections explain the predictor, controller, and simulators in detail, successively.

\subsection{Price Predictor}
\label{sec:predPrice}
Since stablecoins have significantly lower fluctuation than cryptocurrencies \citep{lyons2023stable}, we apply a traditional and robust time series model, Autoregressive Integrated Moving Average (ARIMA) proposed by \citet{box1970ARIMAHolden_Day}, for price prediction. At the beginning of the time interval $(t-1, t]$, we retrain the ARIMA model using the latest $w$ data, which is denoted by $Y_t=\{p_{t-w},p_{t-w+1},...,p_{t-1}\}$. The mathematical expression of ARIMA($m_p,m_d,m_q$), following \citet{brockwell2016arimaSpringer}, is

\begin{equation} \label{arima}
 (1 - \sum_{i=1}^{m_p} \theta_i L^i) (1 - L)^{m_d}p_t = (1 + \sum_{j=1}^{m_q} \theta_j L^j) \varepsilon_t,
\end{equation}
where $m_p,m_d,m_q$ are pre-defined parameters, $\theta_i$ and $\theta_j$ are learned coefficients of autoregressive and moving average components, respectively, $L$ is the lag operator, i.e. $Lp_t=p_{t-1}$, and $\varepsilon_t$ represents forecast error term. The parameters $\theta_i$ and $\theta_j$ are obtained from maximum likelihood estimation. 

\begin{figure} [htbp]%% placement specifier
\centering
\includegraphics[trim=1.7cm 4.7cm 2.4cm 4.7cm, clip, width=1.0\textwidth]{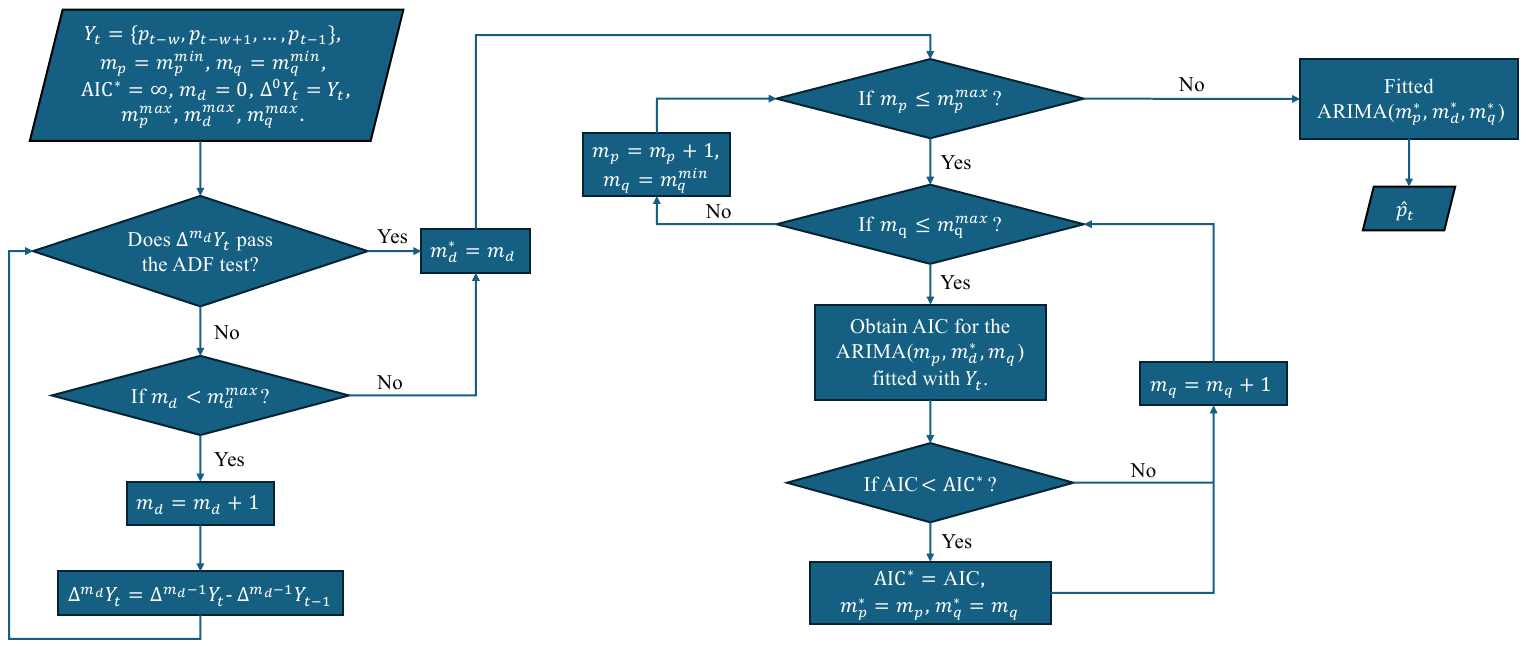}
\caption{Flowchart of the price predictor at the beginning of time interval $(t-1, t]$.}
\label{fig:predictor}
\end{figure}

To determine the order of $m_p,m_d,m_q$ from a pre-defined range: $m_p\in[m_p^{min},m_p^{max}], m_d\in[0,m_d^{max}],\text{ and }m_q\in[m_q^{min},m_q^{max}]$, we apply an automatic selection process following \citet{lyuTSCR} for each period. The optimal differencing order $m^*_d$ is derived by applying the Augmented Dickey-Fuller (ADF) test \citep{dickey1979adfJASA} to the training data across a range of differencing levels ($0 \le m_d \le m^{max}_d$). 
We define the differenced series $\Delta^{m_d} Y_t$ recursively as follows:
$$
\Delta^{m_d} Y_t = \Delta^{m_d-1} Y_t - \Delta^{m_d-1} Y_{t-1}
$$
with the initial series defined as $\Delta^0 Y_t = Y_t$. 
The chosen order is the smallest $m_d$ for which the ADF test yields stationarity of the corresponding data. 
Then we fit the training data across a grid of parameters, fixing the differencing order at $m_d^*$ while varying $m_p \in [m_p^{\min}, m_p^{\max}]$ and $m_q \in [m_q^{\min}, m_q^{\max}]$. The optimal model is selected based on the minimum Akaike Information Criterion (AIC) \citep{akaike1992aic} and is utilized to forecast the price for the subsequent period. 
The prediction process is demonstrated in Figure \ref{fig:predictor}. 

\subsection{Quantity Supplied Controller}
\label{sec:supply}
We apply the Proportional-Integral-Derivative (PID) control law consisting of the sum of proportional, integral, and derivative control actions to adjust the supply of Conttsus. Unlike typical discrete PID control demonstrated in \citet{visioli2006PIDSpringer}, we introduce predicted error at time $t$ to respond promptly to future situations. The adjustment amount of supply for the time interval $(t-1, t]$, $\Delta Q^S_t$, can be obtained through the following controller.
\begin{equation}
\label{eq:deltaSupply}
\Delta Q^S_t=K_p\hat{e}_t^c+K_i\left(\sum_{n=1}^{t-1}e_n^c+\hat{e}_t^c\right)\Delta t+\dfrac{K_d(\hat{e}_t^c-e_{t-1}^c)}{\Delta t},
\end{equation}
where $K_p$, $K_i$, and $K_d$ are positive coefficients, $\Delta t$ is the time difference between two control points, $e_t^c$ is the control error at time $t$ defined in Eq. (\ref{eq:error}), and $\hat{e}_t^c$ denotes the forecast error at time $t$, derived by substituting $\hat{p}_t$ for $p_t$ in Equation (\ref{eq:error}). 
To respond quickly to potential deviations from anchor range $[p^l,p^h]$, we have established a narrower control price range $[p^l_c,p^h_c]$, where $p^l_c>p^l$ and $p^h_c<p^h$. Then we define the control error at time $t$ as 
\begin{equation}
\label{eq:error}
e_t^c = 
\begin{cases}
    p_t-p^h_c, & \text{if } p_t>p^h_c \\
    p_t- p^l_c, & \text{if } p_t<p^l_c \\
    0,   & \text{otherwise.}
\end{cases}
\end{equation}
The quantity supplied at time $t$ is given by 
\begin{equation}\label{eq:supply}
Q^S_t=Q^S_{t-1}+\Delta Q^S_t.
\end{equation}

\subsection{Simulation Methods}\label{sec:simulation}

To verify the effectiveness of the model, we need to conduct simulation experiments to observe price trajectory. 
By assuming market clearing, price at the end of time interval $(t-1, t]$ can be calculated by Eq. (\ref{eq:price}), referring to \citet{akcin2024controlpayIEEE}.
\begin{equation}
\label{eq:price}
p_t=\dfrac{Q^D_{t}}{Q^S_{t}}.
\end{equation}

Quantity demanded is simulated based on the price elasticity of demand, which quantifies the sensitivity of consumer demand to fluctuations in price \citep{mankiw2004elasticityTSW}. 
Once we assume that the price elasticity of the demand $E_d$ remains constant over $[t-1,t]$, we can estimate the demand at time $t$ from the equation

\begin{equation}
\label{eq:demand}
    Q^D_t=Q^D_{t-1}\times\dfrac{1+E_d\cdot(p_t-p_{t-1})/(p_t+p_{t-1})}{1-E_d\cdot(p_t-p_{t-1})/(p_t+p_{t-1})}.
\end{equation}
It is derived from the formula of calculating $E_d$ by the midpoint method. For $E_d\neq1$, substituting Eq. (\ref{eq:demand}) into Eq. (\ref{eq:price}) yields the unique positive root $p_t$ as follows.

\begin{equation}
\label{eq:price_t}
p_t=\frac{p_{t-1}}{2}\left[\sqrt{a^2(1-b)^2+4b}-a(1-b)\right],
\end{equation}
where $a=\frac{1+E_d}{1-E_d}$ and $b=\frac{Q^D_{t-1}}{Q^S_{t}p_{t-1}}$

To better capture the conditions of the real-world, we incorporate random fluctuations in demand. 
The impact of external shocks on demand comprises three cases: a permanent increase of $\alpha$ ($\alpha>0$) times, a permanent decrease of $\beta$ ($0<\beta<1$) times, and a temporary decrease of $\beta$ times followed by gradual recovery. For temporary shocks, the logarithmic disturbance decays geometrically with the persistence parameter \(\rho\) ($0<\rho<1$). At each simulation step, a shock occurs with probability $\eta$, with each type equally likely; overlapping shocks combine multiplicatively. Let $f_j(k)$ be the contribution of the $j$-th shock evaluated $k$ periods following its occurrence. The values of $f_j(k)$ corresponding to the three types of external shocks are $1+\alpha, 1-\beta, (1-\beta)^{\rho^k}$, respectively.
Let $R_t$ denote the cumulative effect of external shocks at time $t$. It can be derived as: $$R_t=\prod_{\tau_j\leq t}f_j(t-\tau_j),$$ where $\tau_j$ represents the time at which the $j$-th external shock is triggered. The initial value $R_0$ is set to 1. Consequently, the demand with random shock can be obtained by $\tilde{Q}^D_t=\frac{R_t}{R_{t-1}}Q^D_t.$ The corresponding price mirrors Eq.~(\ref{eq:price_t}), with $\tilde{Q}^D_{t-1}$ replacing $Q^D_{t-1}$.

\subsection{Stability Analysis}\label{sec:stability}

To determine the conditions for system stability, this subsection applies the $z$-transform \citep{alan2010zTransform} to the baseline model defined by Equations (\ref{eq:deltaSupply}) to (\ref{eq:price_t}). 
To simplify the analysis, we make several assumptions. First, we assume that the predictor is sufficiently accurate so that the prediction error $\varepsilon_t$ in $\hat{p}_{t}= p_{t}+\varepsilon_t$ is negligible. 
Then, we assume that the model can reach equilibrium $(p^*, Q_S^*,Q_D^*)$ at time $t^*$ such that $p_{t}=p^*$, $Q^S_{t^*}=Q_S^*$ and $Q^D_{t^*}=Q_D^* \text{, for }\forall t\geq t^*$ in the absence of disturbances. It is worth noting that the summation part of Eq. (\ref{eq:deltaSupply}) ensures the equilibrium is reached when $e_{t^*}^c=0$, which is $p^*\in[p^l_c,p^h_c]$. Notwithstanding the controller's inactivity within $[p^l_c,p^h_c]$, we can identify the conditions under which the model drives $p_t$ back into this interval from an adjacent external position. 
The following analysis is founded upon the assumption that the system is close to the equilibrium with the differences described as $\delta Q^S_t=Q^S_t-Q_S^*$, $\delta Q^D_t=Q^D_t-Q_D^*$, and $\delta p_t=p_t-p^*$.

The first step is to perform a $z$-transform on the controller. 
By substituting Eq. (\ref{eq:supply}) and $\hat{e}_t^c= e_t^c +O(\varepsilon_t)$ into Eq. (\ref{eq:deltaSupply}), we obtain 
\begin{equation}
\label{eq:controller_all}
Q^S_t-Q^S_{t-1}=K_pe_t^c+K_i\left(\sum_{n=1}^{t}e_n^c\right)\Delta t+\dfrac{K_d(e_t^c-e_{t-1}^c)}{\Delta t}+O(\varepsilon_t).
\end{equation}
Let $E(z)$ and $Q_S(z)$ be the $z$-transform of $e_t^c$ and $\delta Q^S_t$, respectively. Given the identity $\delta Q^S_t-\delta Q^S_{t-1}=Q^S_t-Q^S_{t-1}$ and neglecting the error term, applying the z-transform to Eq. (\ref{eq:controller_all}) results in 
\begin{equation}
\label{eq:supply_z}
Q_S(z)=\frac{1}{1-z^{-1}}\left[K_p+K_i\Delta t\frac{1}{1-z^{-1}}+\frac{K_d}{\Delta t}(1-z^{-1})\right]E(z).
\end{equation}

Then, we attempt to obtain the z-transform of $\delta Q^D_t$, which is represented by $Q_D(z)$. Let $r_t=\frac{p_t-p_{t-1}}{p_t+p_{t-1}},$ Eq. (\ref{eq:demand}) can be expressed as $Q^D_t=Q^D_{t-1}\times\frac{1+E_d\cdot r_t}{1-E_d\cdot r_t}$. By applying the first order Taylor expansion, we have $Q^D_t = Q^D_{t-1}(1+2E_d\cdot r_t)+O(r_t^2)$. After introducing the relationship of $Q^D_t$ and $Q^D_{t-1}$ with $Q_D^*$, we can further obtain $\delta Q^D_t=\delta Q^D_{t-1}+2E_d\cdot r_t\cdot Q_D^* + O(r_t^2+(\delta Q^D_{t-1})^2)$. Since $r_t=\frac{\delta p_t-\delta p_{t-1}}{2p^*}+O((\delta p_t)^2+(\delta p_{t-1})^2)$ , $\delta Q^D_t$ can be represented by $\delta Q^D_{t-1}+\frac{E_dQ_D^*}{p^*}(\delta p_t-\delta p_{t-1})+O((\delta p_t)^2+(\delta p_{t-1})^2+(\delta Q^D_{t-1})^2)$. Therefore, omitting the error term, $Q_D(z)$ can be calculated by 
\begin{equation}
\label{eq:demand_z}
Q_D(z)=\frac{E_dQ_D^*}{p^*}P(z),
\end{equation}
where $P(z)$ is the z-transform of $\delta p_t$. The next step is to find the equation of $P(z)$. 
The total derivative of Eq. (\ref{eq:price}) at equilibrium provides $\delta p_t=\frac{1}{Q_S^*}(\delta Q^D_t-p^*\cdot\delta Q^S_t)$. Therefore, we have
\begin{equation}
\label{eq:price_z}
P(z)=\frac{1}{Q_S^*}Q_D(z)-\frac{p^*}{Q_S^*}Q_S(z).
\end{equation}
Substituting Eq. (\ref{eq:demand_z}) and $p^*=\frac{Q_D^*}{Q_S^*}$ into Eq. (\ref{eq:price_z}), we obtain 
\begin{equation}
\label{eq:supply_z2}
Q_S(z)=\frac{(E_d-1)Q_S^*}{p^*}\cdot P(z).
\end{equation}

To generate the characteristic equation of the model, the remaining thing is to find out the relationship between $E(z)$ and $P(z)$. 
When $p_t>p_c^h$, by Eq. (\ref{eq:error}), $e_t^c=p_t-p_c^h=\delta p_t+p^*-p_c^h$. 
Thus, we have 
\begin{equation}
\label{eq:elasticity_price_z}
E(z)=P(z)+\frac{1}{1-z^{-1}}(p^*-p_c^h).
\end{equation}
By substituting Eq. (\ref{eq:supply_z}) and Eq. (\ref{eq:elasticity_price_z}) into Eq. (\ref{eq:supply_z2}), we obtain 
$$\left[\frac{(E_d-1)Q_S^*}{p^*}-\frac{1}{1-z^{-1}}M(z)\right]P(z)=\left(\frac{1}{1-z^{-1}}\right)^2M(z)(p^*-p_c^h),$$
where $M(z)=K_p+K_i\Delta t\frac{1}{1-z^{-1}}+\frac{K_d}{\Delta t} (1-z^{-1}).$ Therefore, the characteristic equation of the system is
\begin{equation}
\label{eq:characteristic_equation}
\frac{(E_d-1)Q_S^*}{p^*}-\frac{1}{1-z^{-1}}K_p-K_i\Delta t\left(\frac{1}{1-z^{-1}}\right)^2-\frac{K_d}{\Delta t}=0.
\end{equation}
The poles of the system is $z=1$ and the roots of the characteristic equation. Eq. (\ref{eq:characteristic_equation}) can be written as $a_2z^2+a_1z+a_0=0,$ where $a_2=K_p+K_i\Delta t+\frac{K_d}{\Delta t}-\frac{(E_d-1)Q_S^*}{p^*},$ $a_1=-K_p-2\frac{K_d}{\Delta t}+\frac{2(E_d-1)Q_S^*}{p^*},$ and $a_0=\frac{K_d}{\Delta t}-\frac{(E_d-1)Q_S^*}{p^*}.$ Based on the Jury stability criterion \citep{jury1964theory}, a discrete system having the characteristic equation $a_2z^2+a_1z+a_0=0$ (where $a_2>0$) is stable when it meets two specific conditions: $|a_0|<a_2$, and $|a_1|<a_0+a_2$. Consequently, the system can ultimately achieve a stable state as it approaches the control price range $[p^l_c,p^h_c]$, if the following conditions are satisfied: $K_i\Delta t>0$, $K_p+\frac{1}{2}K_i \Delta t>2(C^*-\frac{K_d}{\Delta t})$ and $K_p+K_i\Delta t>max\{0,2(C^*-\frac{K_d}{\Delta t})\}$, where $C^* = \frac{(E_d-1)Q_S^*}{p^*}$. This provides an important reference for the selection of parameters. 

\section{Numerical Experiments}
\label{sec:experiment}

\subsection{Data and Initial Settings}
\label{sec:data}
To start with, we collect historical data of TerraUSD from May 1, 2021, to
April 30, 2022, during which the price of the stablecoin remained relatively stable. Data are downloaded from the CoinGecko platform\footnote{https://www.coingecko.com/}. Note that the original TerraUSD is called TerraClassicUSD on the CoinGecko, to differentiate it from the latest version. Historical data of Ampleforth(AMPL) from January 1, 2022 to December 31, 2022 is also downloaded from CoinGecko. Then, we conduct multiple experiments using historical data from different periods as the initial training set. The first $w$ days of the acquired training data are denoted as $p_{-w+1}, p_{-w+2}, ..., p_0$ and are used as initial historical price values for the simulation process.

We set the anchor range $[p^l,p^h]$ as $[0.995,1.005]$, and the control price range $[p^l_c,p^h_c]$ as $[0.997,1.003]$. Since theoretical arbitrage is free from the risk of loss, market participants are incentivized to pursue opportunities, regardless of how minimal the returns may be. The initial quantity demanded $Q^D_0$ and the initial quantity supplied $Q^S_0$ are both set to $100$. The interval between any two adjacent time stamps $\Delta t$ is 1 day. 

\subsection{Simulation Results}
We perform simulation experiments using the following model parameters. The predictor parameters $m_p^{\min}, m_p^{\max},m_d^{\max},m_q^{\min}, m_q^{\max}$ are given $1$, $2$, $2$, $1$, $2$, respectively. The controller parameters $K_p,K_i,K_d$ are $70,10,20$. The price elasticity of demand $E_d$ is $-1.3$. The initial training set is constructed from data extracted for multiple months of the two separate datasets—starting at the beginning of each month—with the training window size $w$ set to 30, adjusting February to 28. Across all 24 simulation runs, each spanning 150 periods, the simulated prices successfully converge to an equilibrium state in every instance. 

\iffalse
\begin{figure} [htbp]%% placement specifier
\centering
\includegraphics[scale=0.5]{figures/realTrain30_PI2D70_10_20_Dt4_Ed1d3_150.png}
\caption{Simulation results.}
\label{fig:simulationResults1}
\end{figure}
\fi

\begin{figure}[htbp]
    \centering
    \begin{subfigure}[b]{1\textwidth}
        
        \includegraphics[trim=0cm 1.6cm 0cm 0cm, clip, width=1.0\textwidth]{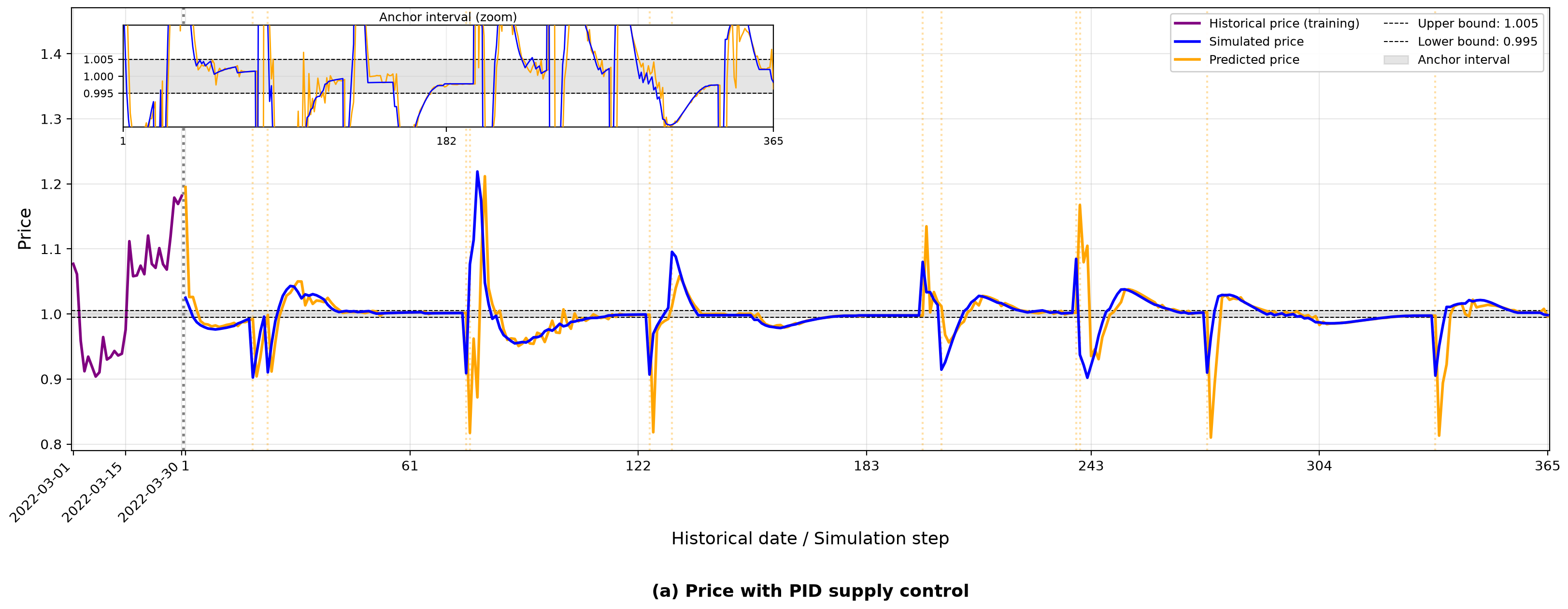}

        \caption{}
        \label{fig:pricesPID}
    \end{subfigure}
    \hfill  % 产生水平间距
    \begin{subfigure}[b]{1\textwidth}
    
        \includegraphics[trim=0cm 1.6cm 0cm 0cm, clip, width=1.0\textwidth]{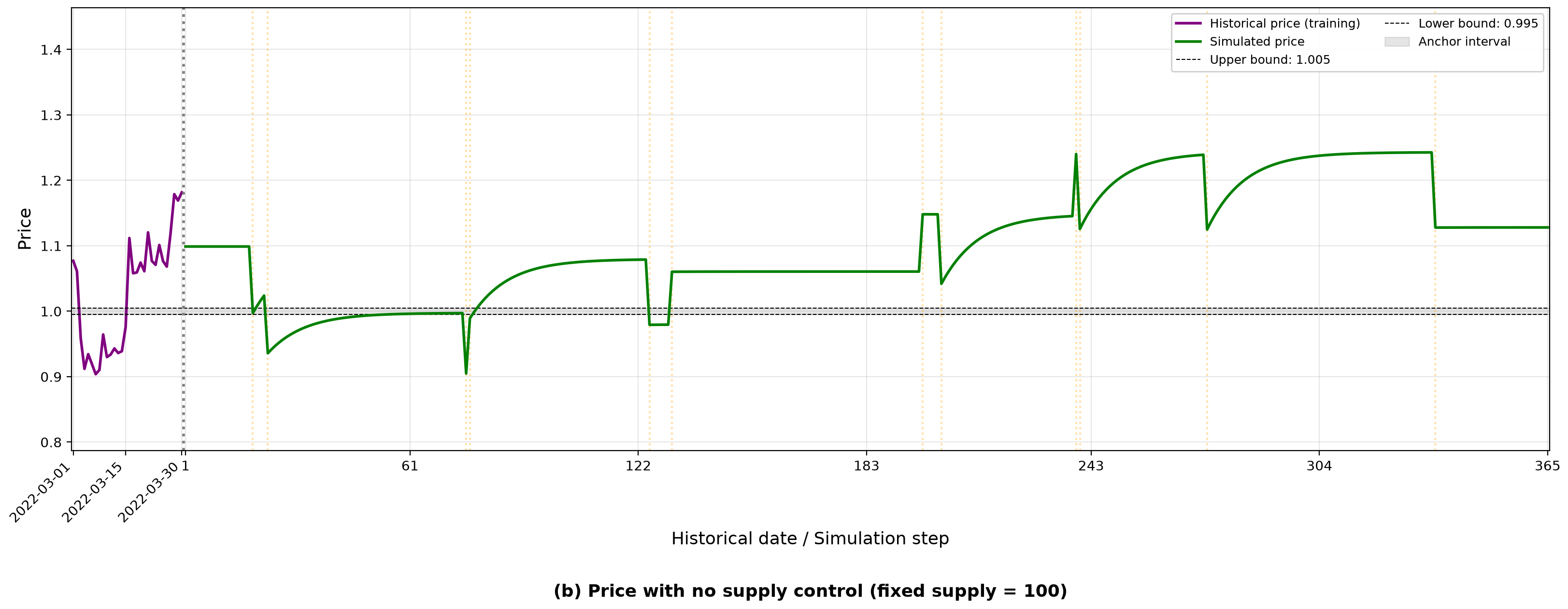}
        \caption{}
        \label{fig:pricesNoController}
    \end{subfigure}
    \hfill  % 产生水平间距
    \begin{subfigure}[b]{1\textwidth}
        
        \includegraphics[trim=0cm 1.6cm 0cm 0cm, clip, width=1.0\textwidth]{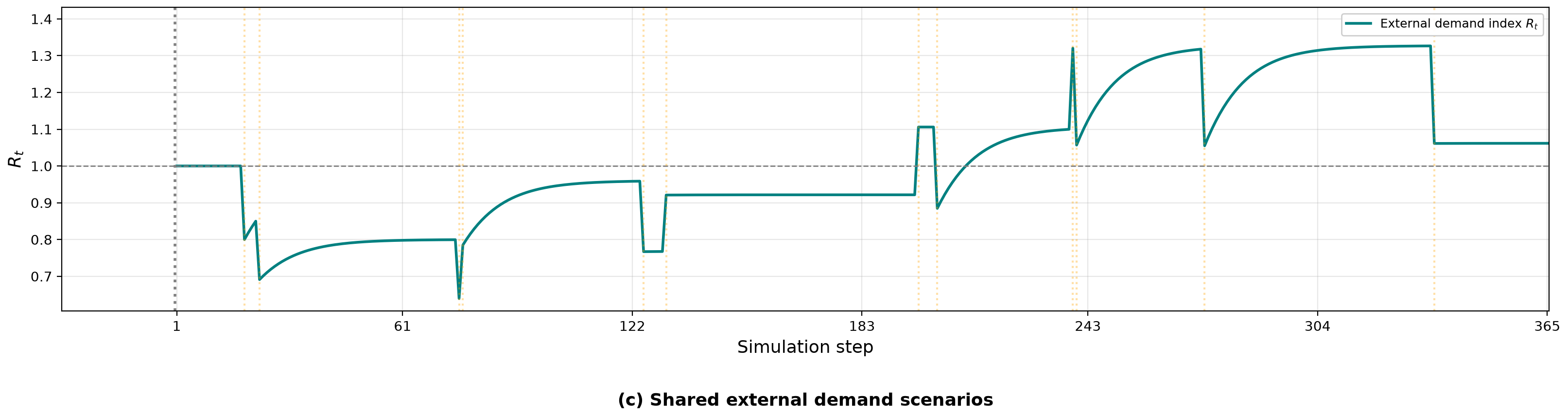}
        \caption{}
        \label{fig:externalShocks}
    \end{subfigure}
    \caption{Simulation outcomes from a 365-day experiment initialized using the March 2022 AMPL dataset for training. The gray vertical dashed line separates historical data from simulation experiments. The yellow vertical dashed lines indicate the moments when external shocks are triggered. (a) Actual historical price (purple), simulated price with PID controller (blue), predicted price (orange), and anchor interval (gray area). (b) Actual historical price (purple), simulated price with no supply controller (green), and anchor interval (gray area). (c) Cumulative effect of external shocks at simulation step $t$, $R_t$.}
    \label{fig:AMPLsimulation}
\end{figure}

Subsequently, we perform experiments that incorporate random external shocks, as detailed in Section \ref{sec:simulation}, with parameters set to $\eta = 0.02$, $\alpha = 0.2$, $\beta = 0.2$, and $\rho = 0.9$. The simulation horizon is extended to 365 days to capture a broader spectrum of potential dynamic trajectories. To evaluate the performance of the PID supply controller, we establish a control group with a constant supply value of 100. Following the previous experimental design, we conduct a total of 24 simulation runs, initialized with each month across the two datasets. Each individual simulation run utilizes the exact same external shock sequence, irrespective of whether the PID controller is active. Across all 24 experiments, the results demonstrate that the PID controller successfully drives the price back to the target range following each disturbance. Conversely, in the absence of a supply controller, the simulated price exhibits pronounced volatility when subjected to external disturbances. For instance, we demonstrate the simulation results using data from the AMPL dataset for March 2022 as the initial training set in Figure \ref{fig:AMPLsimulation}. Figure \ref{fig:externalShocks} records the cumulative external shocks at time $t$. Figures {\ref{fig:pricesPID}} and {\ref{fig:pricesNoController}} illustrate the simulated price dynamics under identical external shocks, comparing trajectories with and without PID supply regulation, respectively. In Figure {\ref{fig:pricesNoController}}, the simulation price without supply controller presents persistent deviations from the anchor range, with a mean absolute error of 0.1040, whereas PID controller reduces this error to 0.0133 and brings the final price to 0.9982. As can be seen from the zoomed-in view of the anchor interval in the top-left corner of Figure {\ref{fig:pricesPID}}, simulation prices regulated by a PID controller are able to return to the anchor range following each external shock. These results indicate that the proposed stablecoin, Conttsus, is able to maintain stability under both ideal conditions and in the face of external shocks, although temporary depegging and overshooting remained.

\section{Conclusions} \label{sec:conclusion}

In this work, a new algorithmic stablecoin, Conttsus, founded on control theory is developed. It first utilizes the traditional time series model ARIMA to predict prices at the next time point. The prediction result is then input into the predictive PID model to adjust the supply volume of the stablecoin in real time. Simulation experiments demonstrate the ability of the stablecoin to maintain price within a pre-defined range. In an ideal environment where prices are influenced solely by supply and demand, the price will eventually converge to a certain value within that range. Even in the presence of exogenous shocks, Conttsus successfully restores prices within the target band.
In future work, we will explore the impact of parameter selection on the results. We will also introduce experiments under more complex or extreme conditions.\\

\noindent
{\bf Acknowledgment:} This research work was supported by 
Hong Kong Research Grants Council under Grant number 17309522.

\bibliographystyle{apalike}
\bibliography{references}

@article{pernice2021cryptocurrency,
  title={Cryptocurrency},
  author={Pernice, Ingolf GA and Scott, Brett},
  journal={Internet policy review},
  volume={10},
  number={2},
  pages={1-10},
  year={2021},
  publisher={Berlin: Alexander von Humboldt Institute for Internet and Society}
}

@book{angeloni202moneyInCrisis,
  author = {Angeloni, Ignazio and Gros, Daniel},
  isbn = {9781009390149},
  title = {Money in Crisis: The Return of Instability and the Myth of Digital Cash},
  year = {2025},
  publisher = {Cambridge University Press}
}

@article{strohmeyer2024stablecoinCategories,
  title={Improving DeFi Mechanisms with Dynamic Games and Optimal Control: A Case Study in Stablecoins},
  author={Strohmeyer, Nicholas and Vishwanath, Sriram and Fridovich-Keil, David},
  journal={arXiv preprint arXiv:2410.21446},
  year={2024}
}

@misc{tether2014usdtWhitepaper,
  title={Tether: Fiat Currencies on The Bitcoin Blockchain},
  author={Tether},
  year={2014},
  url = {https://assets.ctfassets.net/vyse88cgwfbl/5UWgHMvz071t2Cq5yTw5vi/c9798ea8db99311bf90ebe0810938b01/TetherWhitePaper.pdf},
  note = {Accessed: 2026-03-11}
}

@misc{circle2018usdcWhitepaper,
  title={USDC: AN INVESTMENT CRYPTOCURRENCY WITH A PRICE GUARANTEE PLAN},
  author={Circle},
  year={2018},
  url = {https://cryptoactu.com/wp-content/uploads/2018/12/WhitePaper.pdf},
  note = {Accessed: 2026-03-11}
}

@misc{maker2017daiWhitepaper,
  author = {Maker, Team},
  title = {The Dai Stablecoin System Whitepaper},
  year = {2017},
  url = {https://makerdao.
com/whitepaper/DaiDec17WP.pdf},
  note = {Accessed: 2026-03-04}
}

@misc{kuo2019ampleforth,
  title={Ampleforth: A new synthetic commodity},
  author={Kuo, Evan and Iles, Brandon and Cruz, Manny Rincon},
  journal={Ampleforth White Paper},
  year={2019},
  url = {https://www.securities.io/wp-content/uploads/2022/05/Ampleforth.pdf},
  note = {Accessed: 2026-03-09}
}

@misc{kazemian2020fraxV1,
  title={Frax: A fractional-algorithmic stablecoin protocol},
  author={Kazemian, Sam and Huan, Jason and Shomroni, Jonathan and Iyer, Kedar},
  type        = {White Paper},
  year={2020},
  url = {https://github.com/FraxFinance/frax-solidity/blob/master/frax_whitepaper_v1.pdf},
  note = {Accessed: 2026-03-09}
}

@misc{inoescu2020rai,
  title={Rai: A Low Volatility, Trust Minimized Collateral for the Defi Ecosystem},
  author={Inoescu, SC and Soleimani, Ameen},
  year={2020},
  url = {https://resources.cryptocompare.com/asset-management/730/1675951725233.pdf},
  note = {Accessed: 2026-03-10}
}

@inproceedings{zhao2021algorithm,
  title={Understand volatility of algorithmic stablecoin: Modeling, verification and empirical analysis},
  author={Zhao, Wenqi and Li, Hui and Yuan, Yuming},
  booktitle={International Conference on Financial Cryptography and Data Security},
  pages={97-108},
  year={2021},
  organization={Springer}
}

@article{klages2022collateral,
  title={While stability lasts: A stochastic model of noncustodial stablecoins},
  author={Klages-Mundt, Ariah and Minca, Andreea},
  journal={Mathematical Finance},
  volume={32},
  number={4},
  pages={943-981},
  year={2022}
}

@article{calandra2025terraCase,
  title={Learning from Terra-Luna: A simulation-based study on stabilizing algorithmic stablecoins},
  author={Calandra, Federico and Rossi, Francesco P and Fabris, Francesco and Bernardo, Marco},
  journal={Blockchain: Research and Applications},
  pages={100364},
  issn = {2096-7209},
  year={2025},
  publisher={Elsevier}
}

@article{giordano2025cryptocurrencies,
  title={Displacement and complementarity between cryptocurrencies and legacy money},
  author={Giordano, Matteo},
  journal={The Japanese Political Economy},
  volume={51},
  number={1-2},
  pages={80-104},
  year={2025},
  publisher={Taylor \& Francis}
}

@inproceedings{moradi2001predictivePID,
  title={PREDICTIVE PID CONTROL: A new algorithm},
  author={Moradi, Mohammad H and Katebi, MR and Johnson, MA},
  booktitle={IECON'01: The 27th Annual Conference of the IEEE Industrial Electronics Society (Cat. No. 37243)},
  volume={1},
  pages={764-769},
  year={2001},
  organization={IEEE}
}

@book{mankiw2004elasticityTSW,
  author = {Mankiw, N. Gregory},
  year = {2004},
  title = {Principles of microeconomics},
  publisher = {Thomson/South-Western},
  edition = {3rd}
}

@book{box1970ARIMAHolden_Day,
  author={Box, George and Jenkins, Gwilym M},
  year={1970},
  title={Time series analysis: forecasting and control},
  publisher = {Holden-Day}
}

@book{brockwell2016arimaSpringer,
    author = {Brockwell, Peter J and Davis, Richard A},
    title = {Introduction to Time Series and Forecasting},
    publisher = {Springer Nature},
    year = {2016},
    edition = {3rd}
}

@article{samad2017controlSurvey,
  author  = {Samad, Tariq},
  title   = {A Survey on Industry Impact and Challenges Thereof [Technical Activities]},
  journal = {IEEE control systems}, 
  year    = {2017},
  volume  = {37},
  number = {1},
  pages   = {17-18}
}

@book{visioli2006PIDSpringer,
  author={Visioli, Antonio},
  year={2006},
  title={Practical PID control},
  publisher = {Springer}
}

@article{dickey1979adfJASA,
  title={Distribution of the Estimators for Autoregressive Time Series with a Unit Root},
  author={Dickey, David A. and Fuller, Wayne A.},
  journal={Journal of the American Statistical Association},
  volume={74},
  number={366a},
  pages={427–431},
  year={1979}
}

@incollection{akaike1992aic,
  author    = {Akaike, Hirotogu},
  year      = {1992},
  title     = {Information Theory and an Extension of the Maximum Likelihood Principle},
  booktitle = {Breakthroughs in Statistics},
  publisher = {Springer New York},
  pages     = {610-624}
}

@inproceedings{akcin2024controlpayIEEE,
  title={ControlPay: An Adaptive Payment Controller for Blockchain Economies},
  author={Akcin, Oguzhan and Streit, Robert P and Oommen, Benjamin and Vishwanath, Sriram and Chinchali, Sandeep},
  booktitle={2024 IEEE International Conference on Blockchain (Blockchain)},
  pages={316-323},
  year={2024},
  organization={IEEE}
}

@techreport{kereiakes2019terraWhitePaper,
  author      = {Kereiakes, Evan and Kwon, Do and Maggio, Marco Di and Platias, Nicholas},
  title       = {Terra money: Stability and adoption},
  year        = {2019},
  month        = {April},
  institution  = {Terraform Labs},
  type        = {White Paper},
  url = {https://www.allcryptowhitepapers.com/wp-content/uploads/2019/08/Terra_White_paper.pdf},
  note = {Accessed: 2026-03-05}
}

@article{lee2025stablecoinDepegging,
  title={Stablecoin depegging risk prediction},
  author={Lee, Yi-Hsi and Chiu, Yu-Fen and Hsieh, Ming-Hua},
  journal={Pacific-Basin Finance Journal},
  volume={90},
  pages={102640},
  year={2025},
  publisher={Elsevier}
}

@unpublished{lyuTSCR,
  author = {Lyu, Benmeng and Guo, Sini and Zhao, Yuqiao and Ching, Wai-Ki and Siu, Tak-Kuen},
  title  = {Time Series Forecasting Based Online Portfolio Optimization with Risk Control},
  year   = {2026},
  journal = {International Review of Economics and Finance},
  note   = {Manuscript submitted for publication}
}

@book{alan2010zTransform,
author = {Oppenheim, Alan V. and Schafer, Ronald W.},
address = {Upper Saddle River [N.J},
booktitle = {Discrete-time signal processing},
edition = {3rd ed.},
isbn = {9780131988422},
language = {eng},
publisher = {Prentice Hall},
series = {Prentice Hall signal processing series},
title = {Discrete-time signal processing },
year = {2010},
}

@article{jury1964theory,
  title={Theory and Application of the z-Transform Method},
  author={Jury, Eliahu Ibrahim},
  journal={(No Title)},
  year={1964}
}

@article{lyons2023stable,
  title={What keeps stablecoins stable?},
  author={Lyons, Richard K and Viswanath-Natraj, Ganesh},
  journal={Journal of International Money and Finance},
  volume={131},
  pages={102777},
  year={2023},
  publisher={Elsevier}
}
\end{document}